# Distributed Position Localization and Tracking (DPLT) of Malicious Nodes in Cluster Based Mobile Ad hoc Networks (MANET)

SHAKHAKARMI NIRAJ
DHADESUGOOR R. VAMAN
Department of Electrical and Computer Engineering
Prairie View A & M University (Member of Texas A & M University System)
Prairie View, Texas-77446
USA
E-mail: nshakhakarmi@pvamu.edu, drvaman@pvamu.edu, http://cebcom.pvamu.edu

*Abstract:* - In this paper, a robust distributed malicious node detection and precise localization and tracking method is proposed for Cluster based Mobile Ad hoc Network (MANET). Certificate Authority (CA) node is selected as the most stable node among trusted nodes, surrounded by Registration Authority nodes (RAs) in each cluster to generate the Dynamic Demilitarized Zone (DDMZ) to defend CA from probable attackers and mitigate the authentication overhead. The RAs also co-operate with member nodes to detect a target node and determine whether it is malicious or not, by providing the public key certificate and trust value. In addition, Internet Protocol (IP) based Triangulation and multi-lateration method are deployed based on using the average time difference of Time of Arrival (ToA) and Time of Departure (ToD) of the management packets. Triangulation uses three reference nodes which are elected within each cluster based on Best Criterion Function (BCF) to localize each member node inside the cluster in 2D. Multi-lateration is employed to localize the malicious target node in 2D using four neighbor nodes. After localization of two consecutive positions, the target node is continuously localized and tracked by a particular node using the modified real time Position Localization and Tracking (PL&T) algorithm by adaptive beam forming and mapping the energy contours of tracking zone into coverage radii distance. The performance of the proposed scheme demonstrates the significant accuracy in the detection of malicious nodes within each cluster.

## 1. Introduction

Mobile Ad Hoc Networks (MANET) has no fixed infrastructure assets and the nodes are highly mobile. Scalable MANET architectures have been designed in terms of clusters where each cluster having a set of nodes and a cluster head are managed by the cluster head [1]. Also, in MANET architecture, both intra-cluster and inter-cluster multi-hop communications are supported. For a node in a cluster needing to communicate with a node in a different cluster, the multi-hop communications path allows the information to go to the cluster head and then use inter-cluster communications across multiple cluster heads to allow the information to reach the destination node in a remote cluster. The detection of malicious node is a difficult problem. It requires continuous localization and tracking of nodes in a dynamically varying resource constrained network topology. It is critical that no single point failure can occur in MANET. The single point failure issue in inter cluster communication has been addressed by Distributed Certificate Authority approach developing Dynamic Demilitarized Zone (DDMZ) around Certificate Authority (CA). One of the stable trusted nodes is elected as CA in each cluster. DDMZ is formed by assigning Registration Authorities (RAs) in each sector of a cluster as the primary security level to prevent potential attacks to CA [2], [3], [4]. This makes robust filtration of malicious nodes by RAs throughout the cluster when requests come for CA and monitor the node behaviour.

For optimal resource utilization during malicious node detection, RAs are selected from the non confidential community depending upon the Optimum Criteria Function (OCF) and location.

N. Shakhakarmi, Research Assistant and D. R. Vaman, TI Endowed Chair. They are with the ARO Center for Battlefield Communications, ECE Department, Prairie View A&M University, Prairie View, Texas 77446, USA.

OCF is the pseudo selection criteria to mitigate selfish behavior of RA nodes, spurs are assigned to inspire nodes to be RA for DDMZ and RA's locations are determined by steered directional antennas. Distributed RAs filter the extreme number of attackers independently and reduce the network traffic drastically for monitoring. Furthermore, DDMZ has uniform coverage throughout the cluster as RAs are assigned in six sectors of a cluster to detect the adversary attackers or false alarms. In Intra cluster, these malicious nodes are detected using RAs as introducers which provide trust value and public key certificates of target node to requesting node. Depending upon the trust value as well as public key, requesting node can detect whether the target is malicious or not [5], [6], [7], [8]. In Inter cluster communications, when malicious nodes send requests to RA for authentication to access CA, both RA and CA observe their past behaviour, public key and trust value to detect malicious intention of nodes.

Node localization is achieved using three reference nodes which are elected based upon Best Criteria Function (BCF) in each cluster and they are used for triangulation method. Position localization of each member node is performed by using the average time difference of Time of Arrival (ToA) and Time of Departure (ToD) of the management packets exchanged with the reference nodes in each cluster using steered directional antennas. Each reference node computes the range of the target node based on the difference between the computed range and the threshold value. If the difference is lesser than or equal to the threshold, then the average of the three range values is chosen as the range of the node from a given reference. Otherwise, the process is repeated. This increases the processing delay as long as the process is repeated.

The above triangulation method is deployed to locate each node inside the cluster. This is done dynamically with lower computational complexity to localize each authenticated member node within each cluster to update geographical location [9], [10], [11]. Also, multi-lateration method using four or more nodes based on the average time difference of Time of Arrival (ToA) and Time of Departure (ToD) of the management packets is deployed to localize the malicious node more accurately than tri-lateration method. This is executed as soon as, RAs broadcast about the malicious node to neighbors using directional antenna. It is more secure and robust using adaptive beam forming so that the malicious node cannot hide, collude or disguise itself even in the worst case of false data triggering and hiding problems.

Some researchers have shown the use of GPS based radio tracking. This technique adds significantly large processing delay as each node broadcasts their GPS location information using Clear To Send (CTS) signal for small time interval because all nodes are either busy in transmitting or receiving data in the most of time. The delay will be significant even if each cluster has limited number of nodes. However, it has been shown that when using real time Position, Location, and Tracking (PL&T) algorithm based on triangulation, the processing delay is significantly reduced since it depends on the ToD and ToA information to compute the range and does not use CTS protocol [12], [13], [14]. The proposed technique uses geometrical application to track the malicious or desired node based on its two previous position information achieved using triangulation based ranging to minimize the error in the detection of malicious node. The narrow tracking zone over the malicious node is created using an adaptive beam forming while energy contours of the zone are mapped into radii distance in a particular trajectory directed by its latest two positions and further improves the accuracy of tracking and minimize the error of detecting malicious node.

# 2. Problem & Proposed Solution

The problem is to detect the presence of malicious nodes within each cluster with high detection accuracy. The proposed solution creates a narrow tracking zone over the malicious node using adaptive beam forming in a particular trajectory by using its latest two positions and uses triangulation for determining the PL&T of the malicious node in each cluster.

The proposed solution includes the following aspects:

- Specifying and executing a set of procedures using election algorithms
- Specifying a method of detection of malicious nodes
- Designing algorithms for localization and tracking of malicious nodes
- Design of Multi-lateration technique

## 2.1 Election Algorithms

CA election is done considering node stability and connection degree. RA election is done on the basis of trust, stability, residual energy and connection degree. Reference nodes are elected depending upon distance, stability, residual energy and connection degree. When the Clusters are formed by segregation with the relative speed and connection degree of nodes then it is essential to elect CA, RAs and reference nodes in each cluster considering their requisites. CA election is done considering node stability and connection degree. RA election is done on the basis of trust, stability, residual energy and connection degree. Reference nodes are elected depending upon distance, stability, residual energy and connection degree. These are explained in detail below:

### 2.1.1 Cluster Head Election Algorithm

The Cluster Head Election Algorithm elects a trusted node as the Cluster Head. When more than one trusted node exists in each cluster, it is essential to elect one trusted node as Cluster Head depending upon node stability to increase the cluster's lifespan as well as to provide the stable reference. Each trusted node receives an election packet or message from a given node for verification of trust. When a trusted node, A receives an election packet from a trusted node B and the trust is ensured by using the packet-authentication and integrity checking. The following steps are used for the election of cluster head amongst the trusted nodes:

- If the hop count is ≥ (greater or equal) to the cluster size, then the node cannot compete for Cluster Head and dropped from the election.
- If the relative mobility of node B is ≤ (lesser than or equal) to that of node A, then node B is elected as the Certificate Authority (CA) and becomes a Cluster Head. Otherwise, node A is elected as the CA and Cluster Head.
- If the relative mobility of node B is equal to that of node A and degree of neighbor of node B is also equal to that of node A, then the node that has the lowest ID is assigned as the CA and the Cluster Head.

### 2.1.2 Registration Authority Election Algorithm

In each Cluster, CA node needs to elect a Registration Authority (RA) for different sectors using sectored directional antennas as shown in Fig.1. Figure 1 illustrates 6 sectors (sector 1 to sector 6). The RAs are chosen based on the Optimum Criteria Function (OCF) as specified below:

$$OCF = Max \sum_{i \in n} F_i \quad (1)$$

$$F = \sum_{i=1}^{4} w_i x_i \quad (2)$$

where, $w_1 > w_2 = w_3 > w_4$ and $\sum_{i=1}^{4} w_i = 1$ (3)

Trust metric ($x_1$) refers the confident level of nodes which is evaluated by the monitoring mechanism according to its contribution in the network like forwarding ratio or others network services. Stability metric ($x_2$) is based on the relative mobility according to the CA node and based on the received signal strength detected at receiving node. Residual energy metric ($x_3$) determines the residual energy level of the nodes which is the private information of a node. Connectivity degree ($x_4$) is the number of links, a node is connected in one hop. A node having higher connectivity degree can cover more nodes for monitoring in the cluster. The RA election algorithm is given as follows.

- CA broadcasts the *start_election* message to its member nodes in sectors from sector-1 to sector-6.
- CA waits for the reply from the member nodes in each sector in different fixed time interval such as $t_1$ interval for sector-1.
- Member nodes which have the Optimum Criteria Function (OCF) inside each sector are assigned as the RA nodes for Dynamic Demilitarized Zone (DDMZ).
- CA sends confirmation to each elected RA in each sector to perform the primary security checking and monitoring role.

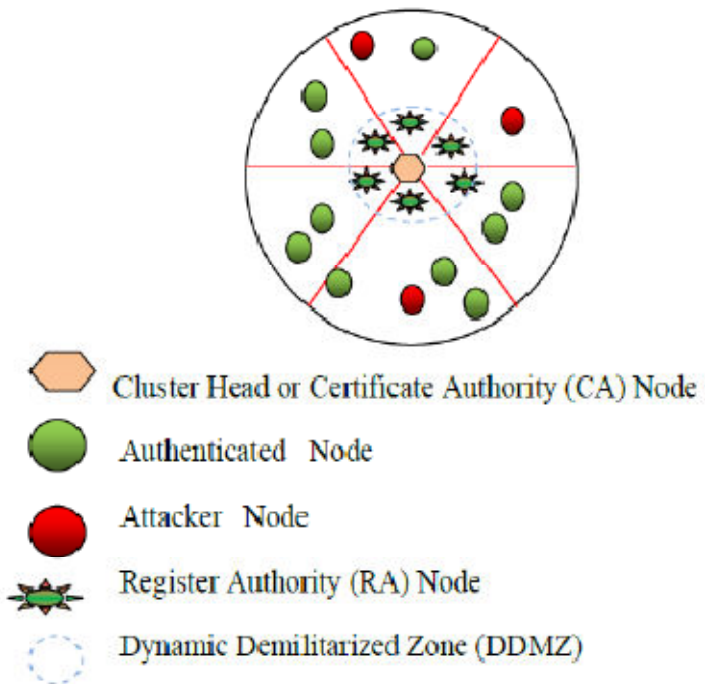

Fig.1 A MANET Cluster model

### 2.1.3 Reference Node Election Algorithm

In each cluster, three reference nodes are to be dynamically elected for triangulation method to locate member nodes. These reference nodes are elected on the basis of the distance from cluster head, stability, residual energy and connectivity degree. Three reference nodes are chosen based on the Best Criteria Function (BCF) as specified below:

$$BCF = Max \sum_{k \in n} F_k \quad (4)$$

$$F_k = \sum_{k=1}^{4} v_k y_k \quad (5)$$

where, $v_1 > v_2 = v_3 > v_4$ and $\sum_{k=1}^{4} v_k = 1$ (6)

Distance metric ($y_1$) is the relative distance taken with respect to cluster head. Stability metric ($y_2$) is based on the relative mobility to the cluster head node and based on the received signal strength detected at receiving node. Residual energy metric ($y_2$) determines the residual energy level of the nodes which is the anonymous information of a node. Connectivity degree ($y_3$) is the number of links, connected in one hop which covers up more nodes to provide reference in the cluster. The reference node election algorithm is given as follows:

- In each cluster, each node computes Best Criteria Function (BCF) and compare with the threshold value. If a node has BCF greater than or equal to the threshold value then it is elected as a candidate, otherwise it cannot be a candidate.
- The elected node broadcasts its BCF within the cluster and a local timer is initialized upon broadcast. The value of this timer is defined so that the node can receive BCF values from each other elected nodes within each cluster before the timer expiration.
- Three nodes having highest BCF values and almost equidistant from each other, are assigned as the reference nodes. This will provide appropriate geometry for tracking of malicious nodes.

Upon election as Reference nodes, the nodes use the Angle of Arrival (AOA), the average time difference of Time of Arrival (ToA) and Time of Departure (ToD) of management packets exchanged between them to compute their PL&T. Thus, the three Reference Nodes have knowledge of each other's PL&T. The Reference Nodes are ready to perform triangulation to track other nodes and identify malicious nodes within each cluster.

## 2.2 Malicious Node Detection

Dynamic Demilitarized Zone (DDMZ) is developed by RA in each sector in each cluster one hop away, surrounding CA to protect from malicious nodes. The major objective of DDMZ is to protect CA against false triggering, Denial of Service (DoS) attack and to monitor the activities of member nodes. In other words, RAs are assigned to avoid the single point of failure at CA in cluster by filtering traffics from authenticated member nodes. RAs provide secure and distributed authentication service by secured public key authentication service to prevent nodes from obtaining false public keys of others when there are malicious nodes in the networks.

Malicious node is detected by verification of node ID, Public key and trust value by deploying some partially trusted nodes. Thus, RAs become partially trusted nodes which perform traffic monitoring and filtering in each sector, at one hop distance with CA and member nodes so that any malicious or misbehaved node can be detected and informed to neighbors.

RAs are distributed to sign public key certificates in each sector of a cluster for authentication. Let us consider 's' be the node requesting for public key of a target node 't' then, node 's' request for public key certificates signed by RA. When node 's' and 't' are in different sectors in a cluster then their RAs are as introducers using the public key certifications. The introducers '$ra_1$', '$ra_2$', ..,'$ra_n$', send reply requesting node 's' with the public key and the trust value of target node 't' upon request. Node s computes the trust value of 't' by using the trust value from '$ra_1$', '$ra_2$', ..,'$ra_n$' and each reply must be signed by the concerned introducer with its private key for corroboration. The algorithm of malicious node detection is given as follows:

- Requesting node's' can achieve the public key certificates and trust value of the target node 't'. First of all, 's' node requests for public key and trust value to its sectored RA. The introducer for 's' node is its sectored RA which send reply message 'm'= { $Pk_t, V_{ik,t}$,..} $Sk_{ik}$ where $Pk_t$ is the public key of node 't', $V_{ik,t}$ is the trust value from $i_k$ to 't' and $Sk_{ik}$ is the private key of $i_k$. Sometimes, node 't' may be in different sectors then RAs are introducers represented as $i_k$ an each introducer concatenate their reply message.
- A new trust relationship, $V_{s,ik,t}$ from 's' to 't' via the intermediate node $i_k$ on a single path, is computed as;

$$V_{s,ik,t} = V_{s,ik} \Theta V_{ik,t} = 1-(1-V_{ik,t})^{Vs,ik} \quad (7)$$

When the requesting node 's' achieve the public key and trust value then it checks the trust value above some threshold value or not. If the trust value is lower than the threshold value then target node is detected as malicious node. If the trusted value is above threshold then it allows the voting on the received public key to its neighbor nodes. If the majority of neighbours agree on $Pk_t$ then it is honest node, otherwise; it is detected as the malicious node.

This algorithm deals that a node can easily detect whether its target is malicious or not in Intra Cluster. The time complexity of algorithm is reduced by assigning RAs in each cluster at one hop. Similarly, for Inter Cluster, when the malicious nodes send request to RA for authentication to access CA then both CA and RA observes its behaviour, public key and trust value. Once any one of these confidential factors are found fault then the node is caught as malicious. RA always broadcast about detected malicious node to its neighbours for PL&T.

### 2.3 Malicious Node Localization and Tracking

Triangulation method is deployed after election of three reference nodes to localize each node inside the cluster. The location information of each node is broadcasted with Clear To Send (CTS) signal only for small time interval to other nodes inside the cluster in GPS based radio system. But, it takes significant time delay as the location information is broadcasted after longtime of data transmission or reception. Therefore, the immediate location computation and tracking in dynamic environment is very challenging task which can be performed by using pre-existing location information and geometrical application, termed as modified Position Localization and Tracking (PL&T). Similarly, multi-lateration method based on the average time difference of Time of Arrival (ToA) and Time of Departure (ToD) is used to locate malicious node from its neighbor nodes and later tracked by modified PL&T.

#### 2.3.1 Triangulation method

In triangulation method, each reference node computes the range (or the distance) between the desired node and itself based on average difference between the Time of Arrival (ToA) of the IP management packets at the desired node and the Time of Departure (ToD) of the IP management packets at the reference node. Figure 2 illustrates the method of triangulation using highly intense directional beam and then translate into the corresponding distance as follows:

$$D_{Euclidean} = c * \sum_{i=1}^{n} (ToA_i - ToD_i) / n \qquad (8)$$

where, c is the speed of light $= 3\times10^8$ m/ sec and n is the number of packets for averaging to determine the range (or distance).

The final range is determined as follows:

If the ranges derived by each of the references are $\leq t_o$ the threshold value, the final range is computed as the average of the three ranges; if two of three ranges are $\leq t_o$ then the threshold value, only those two are used for averaging to compute the final range. If each of the three range values is greater than threshold, the process of range determination is repeated to locate the nodes in the cluster. In addition, each reference computes the Angle of Arrival (AoA) to determine the specific direction.

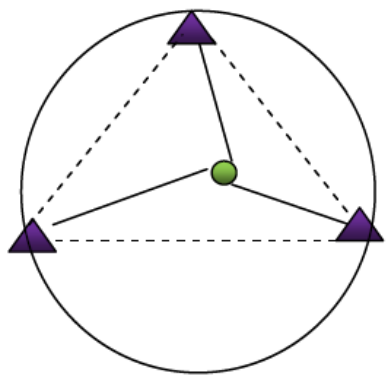

Fig. 2 Illustration of Triangulation in a cluster

#### 2.3.2 Modified Position Localization and Tracking

Transmission of ToD and ToA is often difficult in MANET as nodes are occupying the bandwidth for exchange of user information. One of the ways, the ToD and ToA can be exchanged is through the exchange of management packets containing the information of ToD and ToA of all the packets used for ranging. This will incur use of network bandwidth for tracking which is an overhead function with respect to the user information exchange. The other possibility is the determination of tracking zone by adaptive beam formation so that a desired node can be tracked anywhere before the range information is obtained using triangulation. The adaptive beam formation is used in this effort to develop a modified PL&T operation. Thus, in this method, triangulation is not performed until the zone of a desired node for tracking is developed using the adaptive beam formation. The tracking zone formation is developed by the transmitter using the previous two locations achieved from the triangulation. We assume all the nodes move in a forward direction (within a specific angle) as determined by the previous two locations. We initially draw a straight line and draw a circle

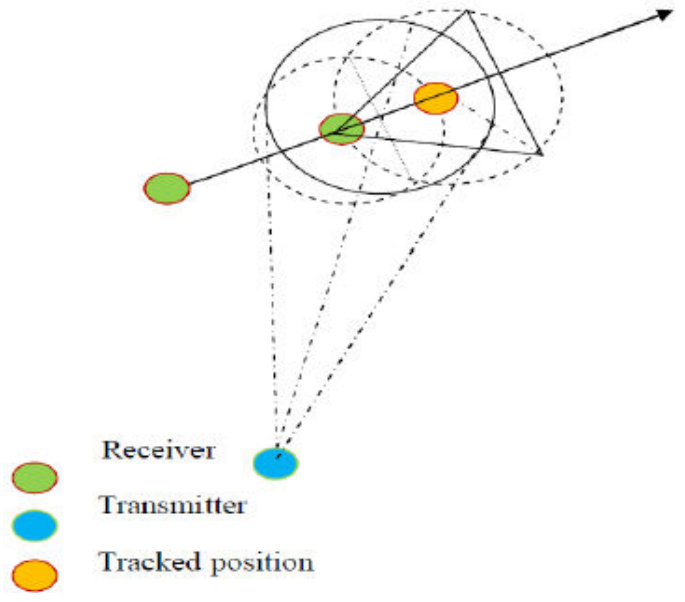

Fig.3 Transmitter deploying PL&T over Receiver (energy contour) at the available latest position as the center. We continue to draw the energy contour by moving forward in straight line equal to radius of the circle and then determine whether the desired node can be located using the directional beams of the two references. The ratio of the coverage radii distance between energy contours of $(n + 1)$ and $n$ is maintained to be$\sqrt{(n + 1)}$, so that the the adjacent contours cover the same area. Once the received signal energy is received from the target node inside the zone, the coverage radii distance is computed. Figure 3 illustrates the adaptive beam formation and determining the coverage zone. Figure 4 illustrates the energy contours where the direction of the node from the previous locations determine the exact location on the particular radii contour.

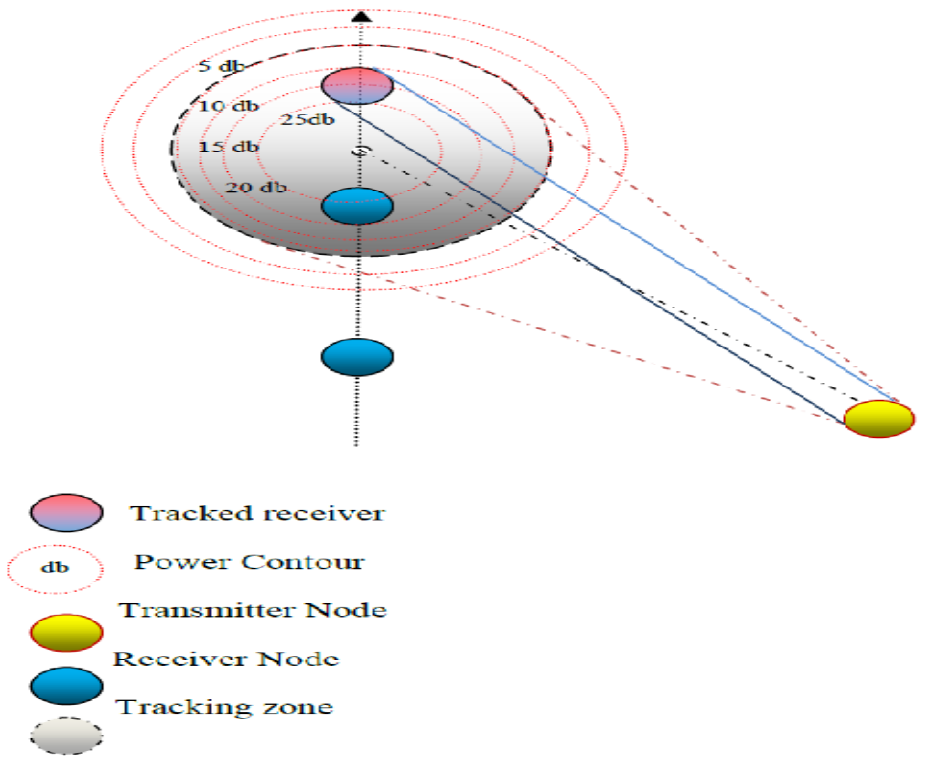

Fig. 4 Illustration of Energy contours

Multiple transmitters can track the desired receivers using above modified PL&T depending upon the latest two location information and labelling energy contours into the coverage radii distance as illustrated in Fig. 5. The same process is deployed by multiple receivers to track the desired transmitters. Then all nodes are familiar with the location of other neighbour nodes inside the same cluster through multi-hop chaining, and they can even track each other as per requirement in the mobile environment.

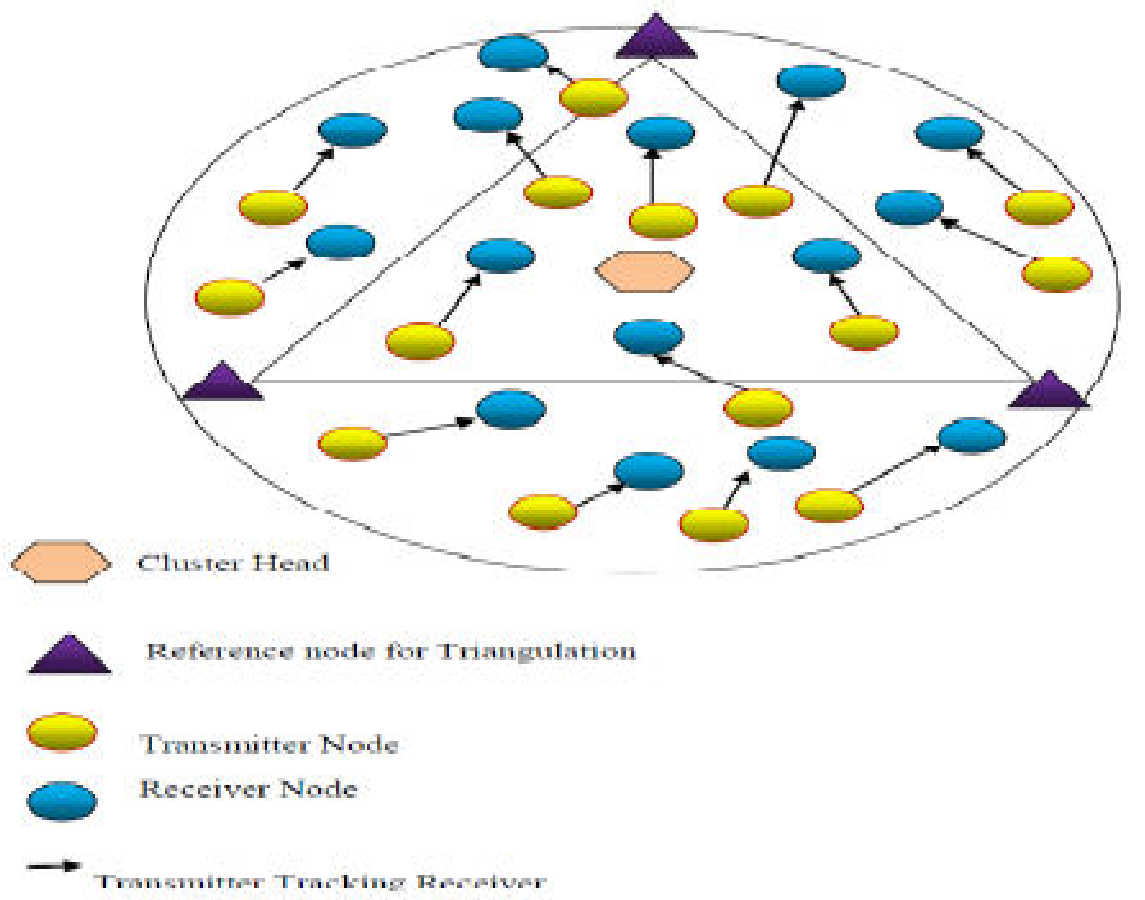

Fig. 5 Multiple Transmitters Tracking Receivers

### 2.3.3 Multi-lateration Method

In proposed multi-lateration method, the PL&T of a malicious node is computed using known (from traingulation) four or more nodes deploying the nearest node to the malicious (target) node. It should be assumed that the malicious node will be attempting to communicate with other nodes within the cluster and therefore, it will be participating in the multilateration process which is performed using the average time difference of ToA and ToD of IP packets. The multi-lateration is illustrated in Fig. 6.

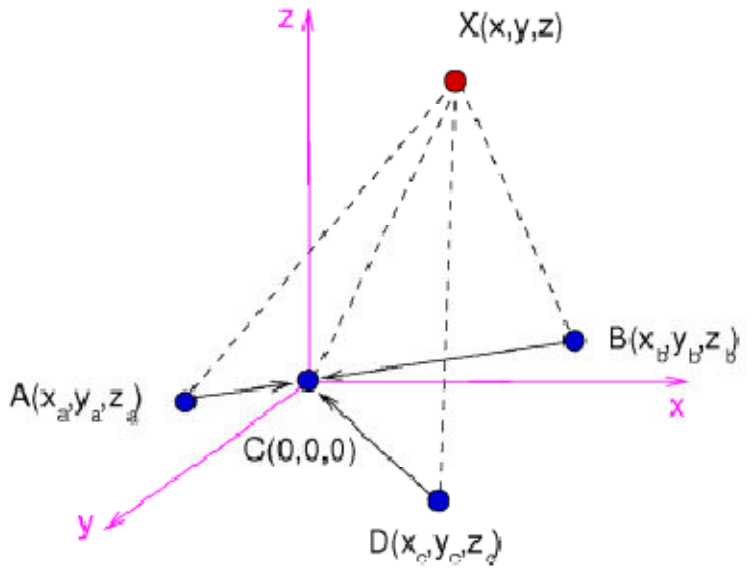

Fig. 6 Multi-lateration using Four Reference nodes

Multi-lateration is precise and robust than trilateration and triangulation as it derives PL&T based on the optimal reference out of three references. In addition, malicious node is exactly localized as it could not trigger any false location data by replaying ToA and ToD packets to four references simultaneously. For four reference nodes $A(x_a, y_a, z_a)$, $B(x_b, y_b, z_b)$, $C(x_c, y_c, z_c)$, and $D(x_d, y_d, z_d)$, are chosen and one of them is assumed to be origin $C(x_c, y_c, z_c) = C(0,0,0)$ and the distances from A,B and D to origin C. Then, the ranges or distances from these nodes to the malicious node X(x,y,z) are determined as follows:

$$T_{m,n} = c * \sum_{p=1}^{q} (ToA_{n,m} - ToD_{m.n}) / q \qquad (9)$$

$$T_{m,C} = c * \sum_{p=1}^{q} (ToA_{C,m} - ToD_{m.C}) / q \qquad (10)$$

$$T_{C,n} = c * \sum_{p=1}^{q} (ToA_{n,C} - ToD_{C,n}) / q \qquad (11)$$

$$D(m,n) = s * \left( T_{m,n} + T_{m,C} - T_{C,n} \right) \qquad (12)$$

where, q is the number of packets, s is the speed of propagation of packet (light), m and n are variables representing sender and receiver such as nodes A,B D and X, and D(m,n) is the derived distance between sender m and receiver n using reference node C.

When the malicious node sends requests to RA for registration to get access to CA, the RA verifies whether it is legitimate member node or not by checking its public key, timestamp, and the trust value. If the node is deemed as not legitimate node (or a member node), the RA multicasts (securely) the information that the node is deemed as malicious node to all other authenticated neighbours in the same sector. Upon receiving this broadcast message, a set of authenticated neighbour nodes (nearest to the deemed malicious node) are selected for multi-lateration localization by the cluster head for determining the PL&T of the malicious node. Fig.7 illustrates the multi-lateration process to detect the PL&T of the malicious node. It is possible that the malicious node may manipulate the process of ToD and ToA deliberately. It can also hide itself using directional antenna and change signal characteristics. However, the neighbouring references can keep track of the zone where malicious node exists and continue to find the PL&T within the zone. Fig. 7 illustrates the multi-lateration process for PL&T determination of the malicious node. Fig. 8 illustrates the multi-lateration process for malicious node in both intra-cluster and inter-cluster multi-hop path connectivity. This will prevent malicious node to behave like a cluster head.

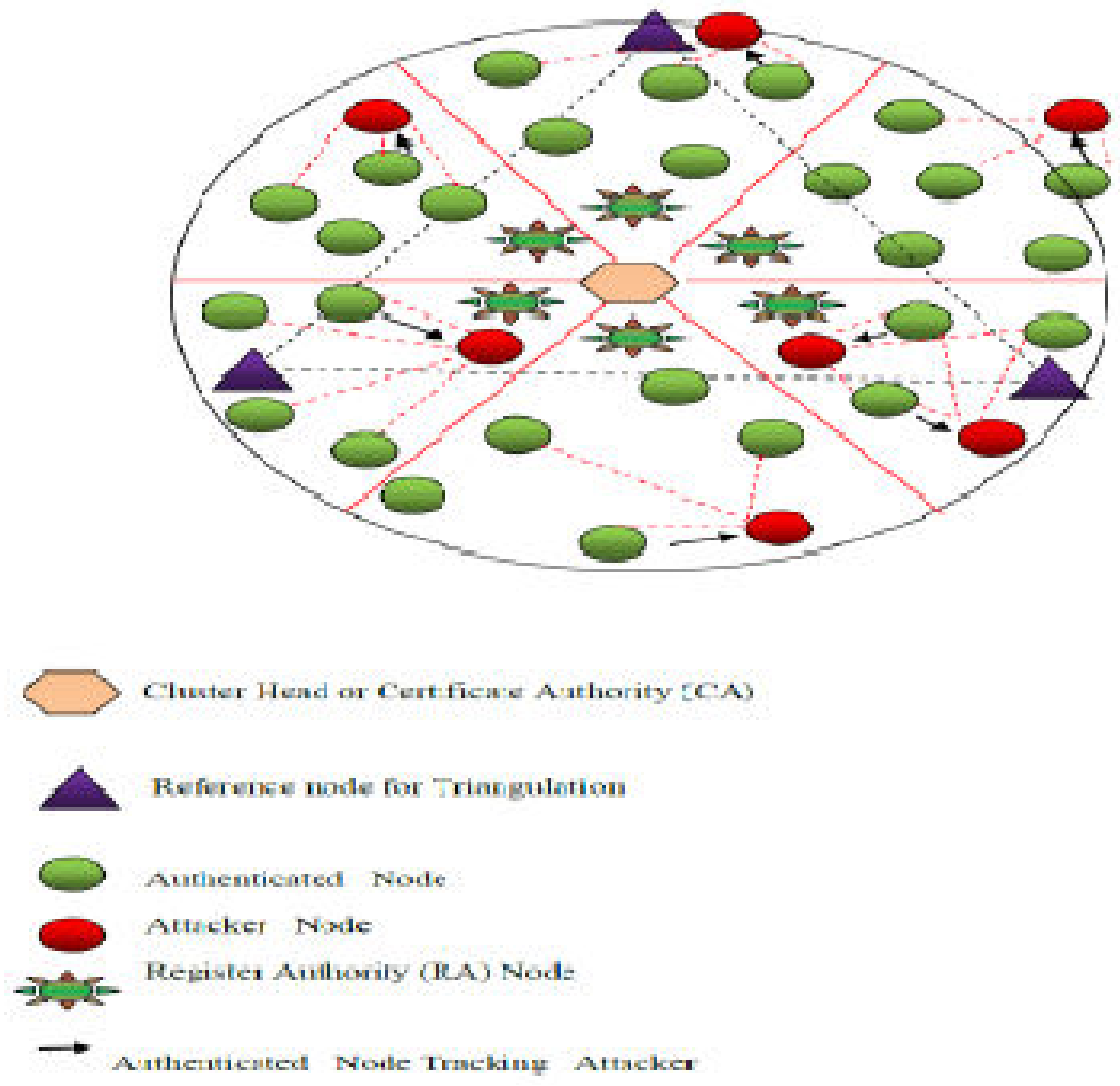

Fig. 7 Authenticated Nodes Tracking Attackers in Cluster based MANET

In inter-cluster, there is a problem of node localization and tracking when some nodes fall in between two clusters which cannot be accessed by either RA or CA or both. In other words, they are out of range nodes. In such circumstances, authenticated nodes which are localized by nearby reference nodes are used to determine the PL&T of the out of range nodes (including malicious nodes).

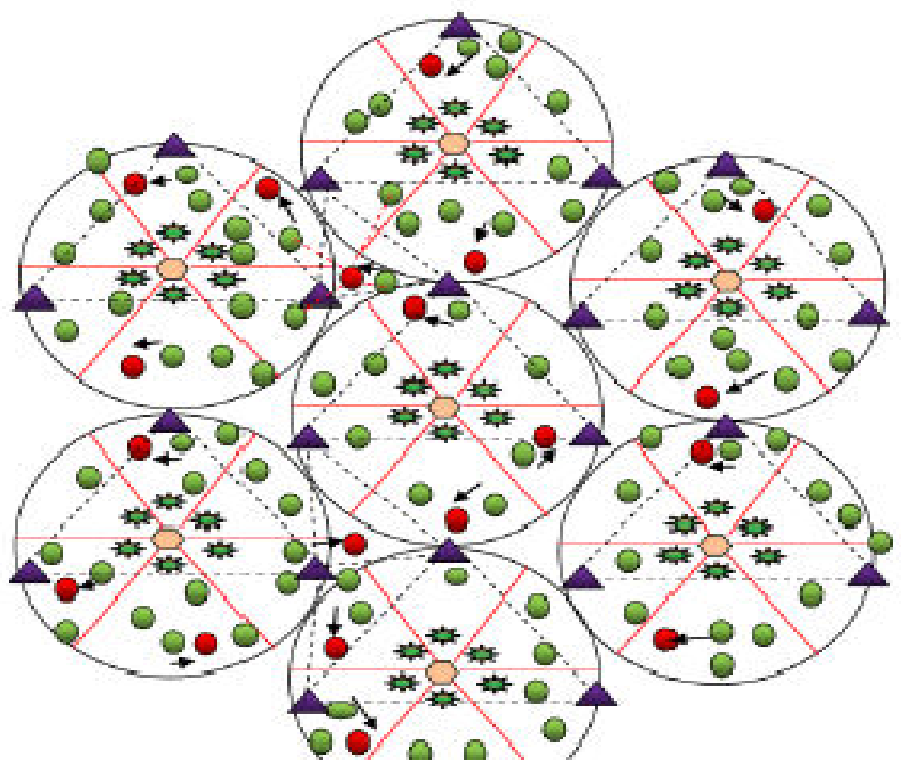

Fig. 8 Authenticated Nodes Tracking Attackers in Intra and Inter Clusters MANET

## 3. Simulation & Performance Results

Simulation is performed using 80 nodes in each cluster with coverage area 700 X 700 square meters, transmission range 300 meters and random way point mobility. First of all, cluster is formed by nodes having similar mobility, residual energy and degree of connection. Then, the Cluster Head is elected from a set of trusted nodes depending upon the stability and degree of connection to assign as Certificate Authority (CA) using CA Election Algorithm. Form simulation results, the dropped nodes CA election and CA candidates are found increasing, however elected node is always between three and eight in the simulation period of 600 seconds which refers the consistency of the algorithm and is shown in Fig. 9.

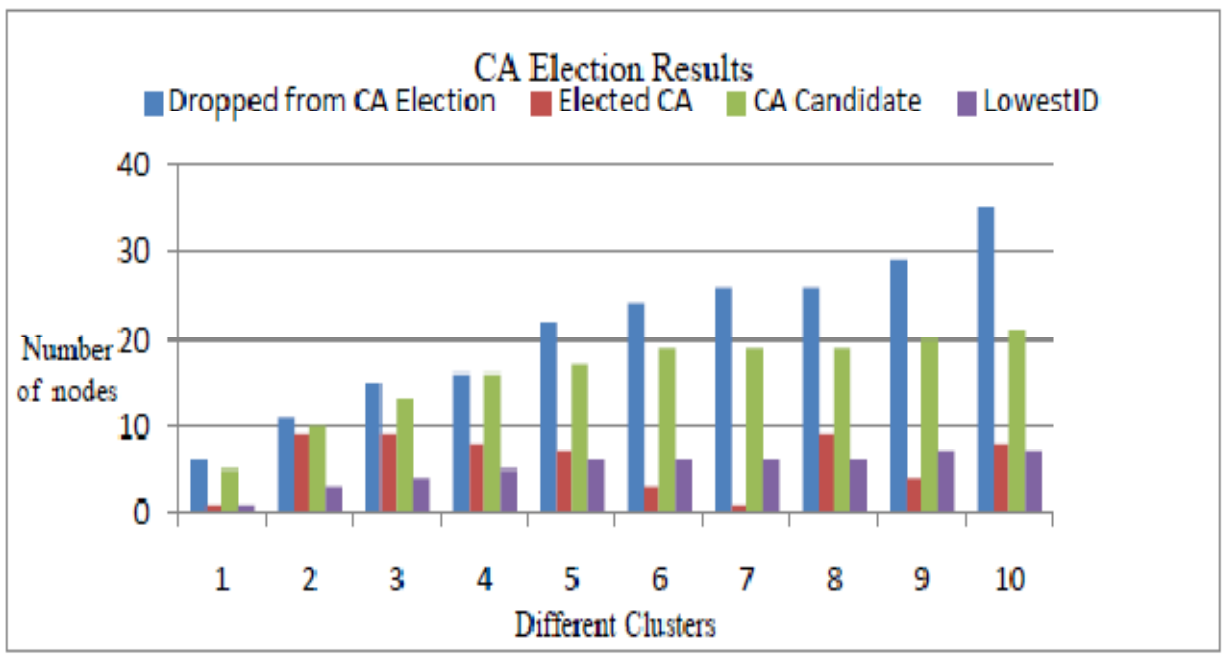

Fig. 9 Certificate Authority Election

Registration nodes (RA) prevent the single point of failure by developing Dynamic Demilitarized Zone (DDZ) around CA. These RAs are elected in each sector of a cluster depending upon Optimum Criteria Function (OCF) consisting trust, stability, residual energy and connectivity degree. The major responsibility of distributed RA architecture is to detect the malicious node precisely in a particular sector and broadcast to the neighbors for localization and tracking. The RA election results that the worst value of OCF is 0.47 in first sector of sixth cluster and the best value as 1 in each cluster while using weighs as $w_1$=0.46, $w_2$ =0.22, $w_3$=0.22 and $w_4$=0.1 as shown in Fig. 10.
Reference nodes election is run to select the three reference nodes in each cluster so that each node inside that cluster can be localized by Triangulation method. Reference nodes are selected on the basis of Best Criteria Function (BCF) which consists of the distance from cluster head, stability, residual energy and connectivity degree. From simulation, BCF is found between 0.96 to1 and candidates for reference node are increasing in different clusters using weighs as $v_1$=0.44, $v_2$=0.23, $v_3$=0.23 and $v_4$=0.1 as shown in Fig. 11.

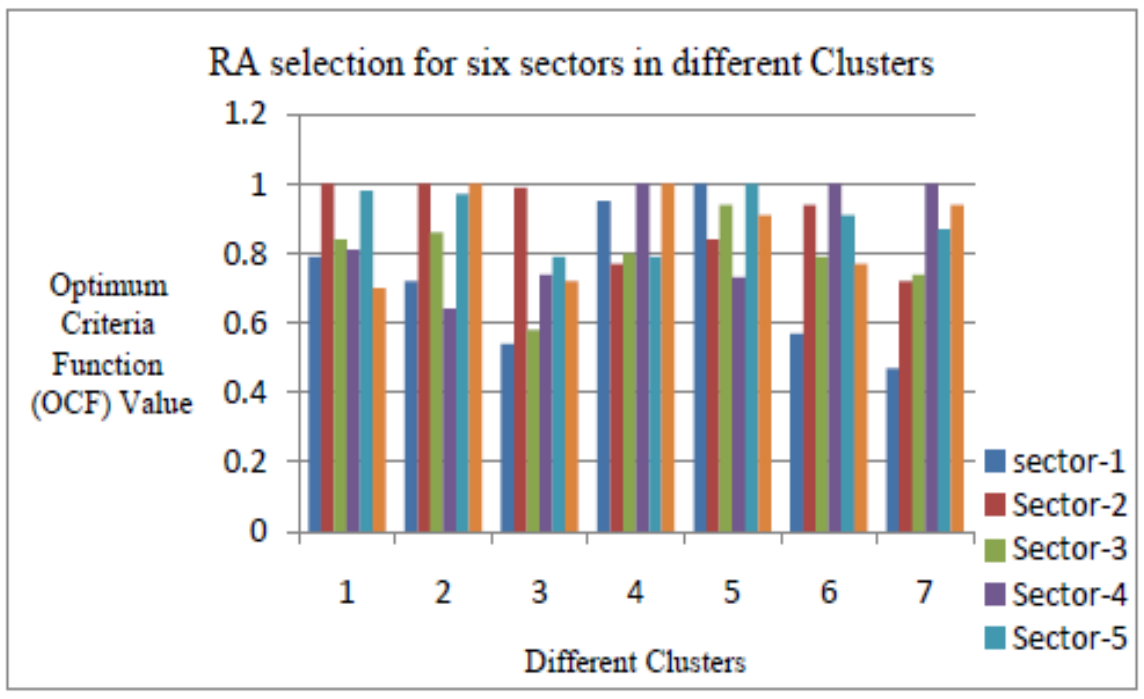

Fig. 10 Registration Authority nodes Election

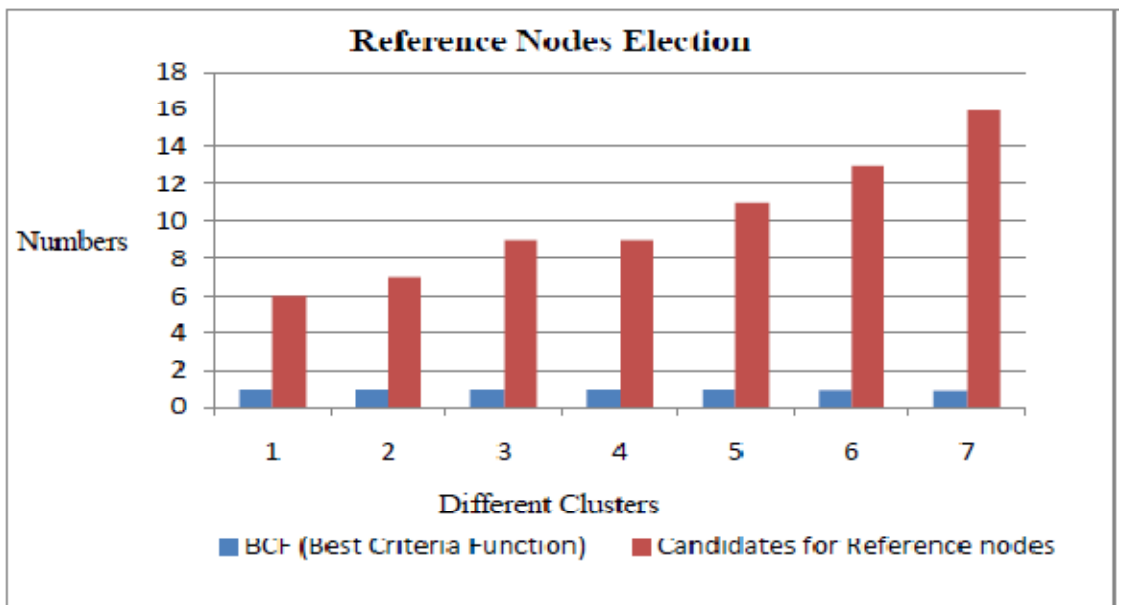

Fig. 11 Reference node Election

The attacks senario in six sectors of a cluster is illustrated for 60 seconds and node behaviour is found that there is always possibility of attack in each sectoral RA as shown in Fig. 12. As the node behaviour is found greater than 0.8 then it is misbehaviour or attack, otherwise it is normal behaviour. From simulation, the misbehaviour or attack is found in sector-1 between 15 to 35 secs, sector-2 between 5 to 15 secs, sector-3 between 35 to 45 secs, sector-4 between 55 to 60, sector-5 between 45 to 50 and sector-6 between 45 and 55.

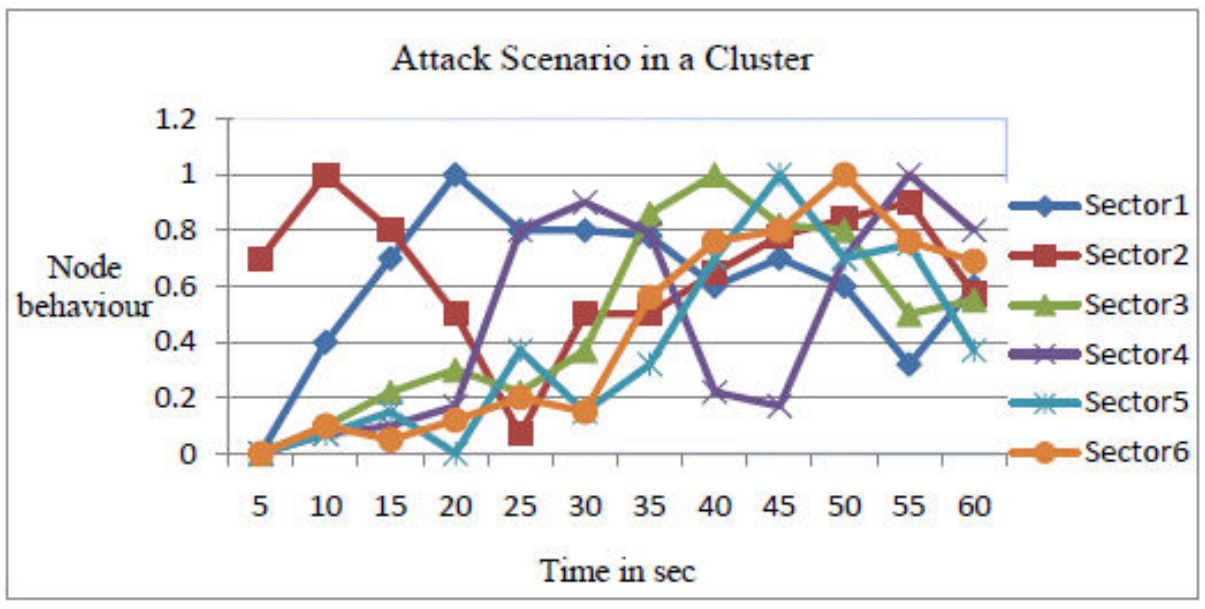

The malicious node detection is executed in all sectors of seven clusters using RAs which provide the public key and trust value of target node to requesting node and later requested node verifies the trust value and the public key. Simulation shows that different numbers of malicious nodes are detected in different clusters such as five malicious nodes in cluster-1, four malicious nodes in cluster-2, eight malicious nodes in cluster-3, eight malicious nodes in cluster-4, eight malicious nodes in cluster-5, seven malicious nodes in cluster-6 and five malicious nodes in cluster-6 as shown in Fig. 13.

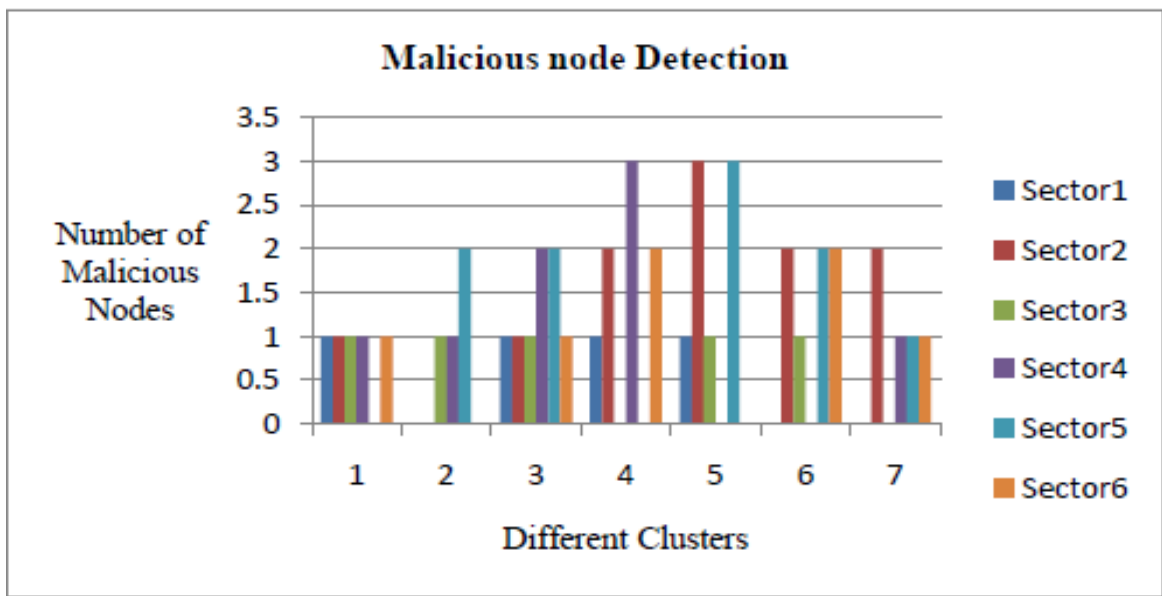

Fig. 13 Trust and PKI based Malicious Detection

Triangulation method is deployed in the cluster by choosing three reference nodes. The average Time Difference Time of Arrival (ToA) and Time of Departure (ToD) is used to compute distances and AoA for angular measurement using 3 packets. If the distances taken from three reference is equal or fall below threshold value 2 m then the three distences are averaged to derive PL&T with the angular information. If only two distances are equal or fall below threshold value 2 m then two distances are averaged to derive PL&T with the angular information. In other cases, the process is repeated. From the simulation, it can be seen that a set of three reference points are found at: [(1,0),(134,13),(213,108)],[(27,88),(43,279),(338,20)],[(82,191),(87,170),(571,4119)],[(66,119),(219,288),(68,203)],[(56,68),(347,105),(107,211)]} as shown in Fig.14.

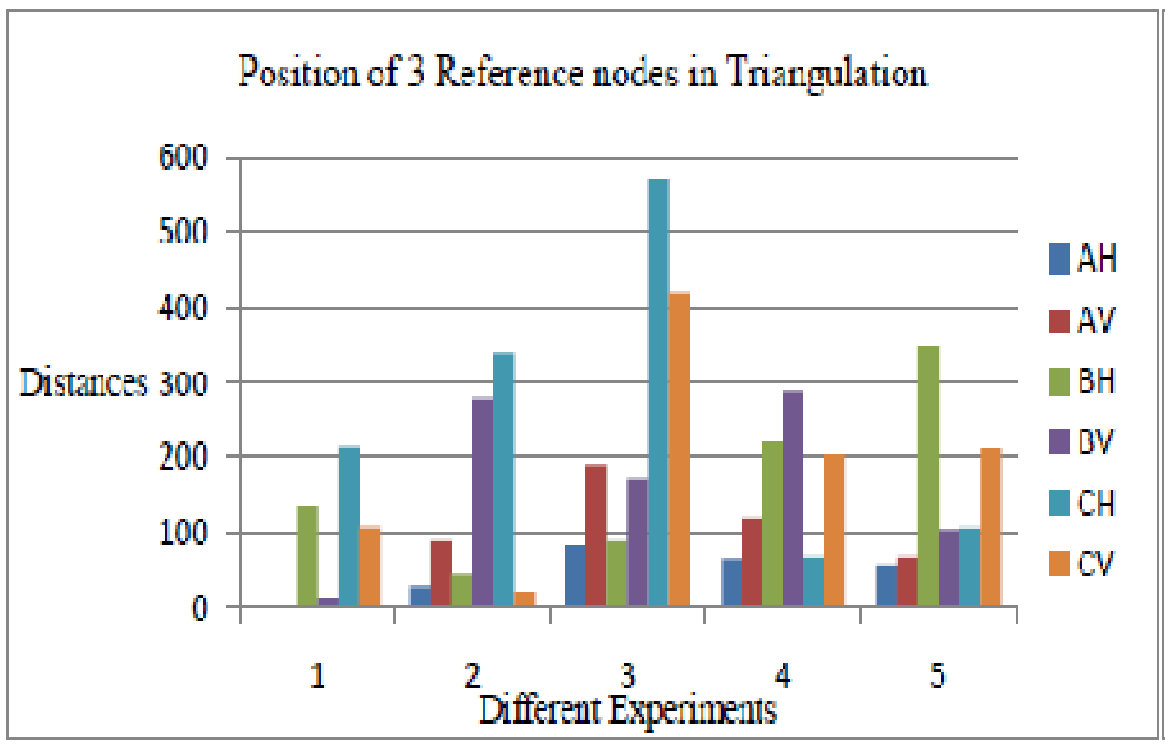

The corresponding distances between above reference points are found{(121,123,237),(303,392,318),(21,544,539),(227,173,84),(293,262,151)} as shown in Fig. 15.

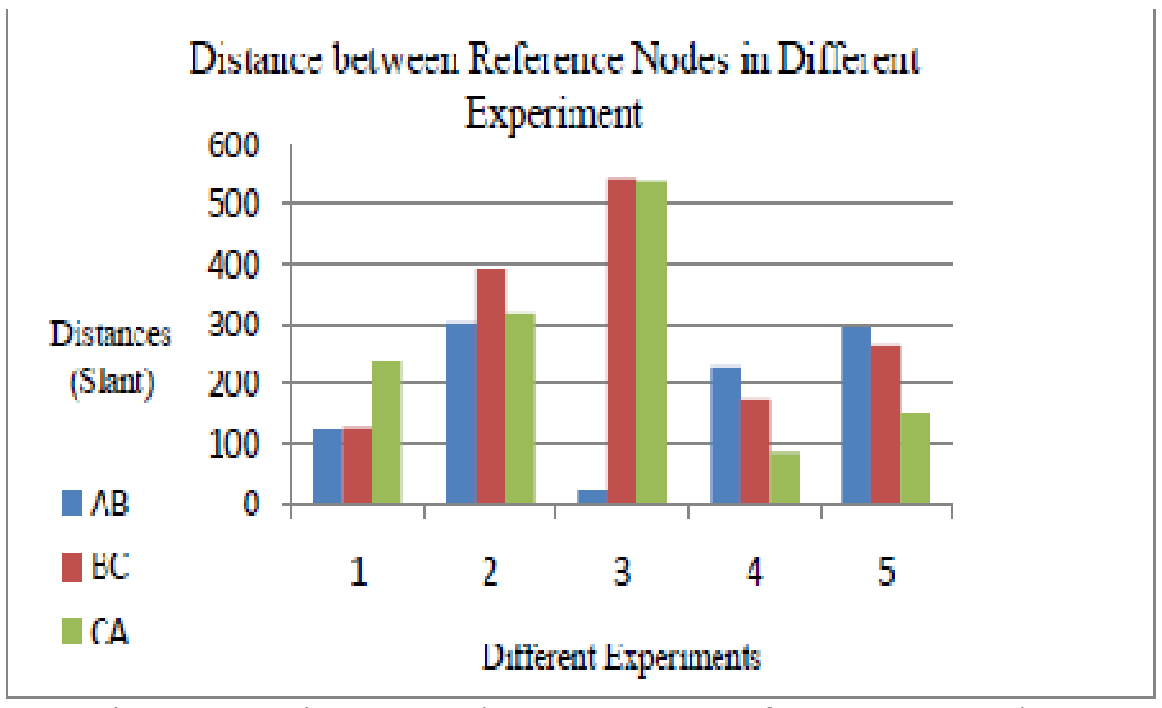

Fig. 15 Distance between Reference nodes

Modified Position Localization and Tracking is executed in our simulation. Tracking authenticated nodes using modified PL&T and the distances are illustrated in Fig. 16. Also from the simulation, while tracking a node at different speed, it is found that the tracked location is directly proportional to the speed. The minimum tracked location is found (205,153) with slant distance 255 at 10m/s and maximum tracked location (463,341) with slant distance 679 at 100m/s as shown in Fig. 17.

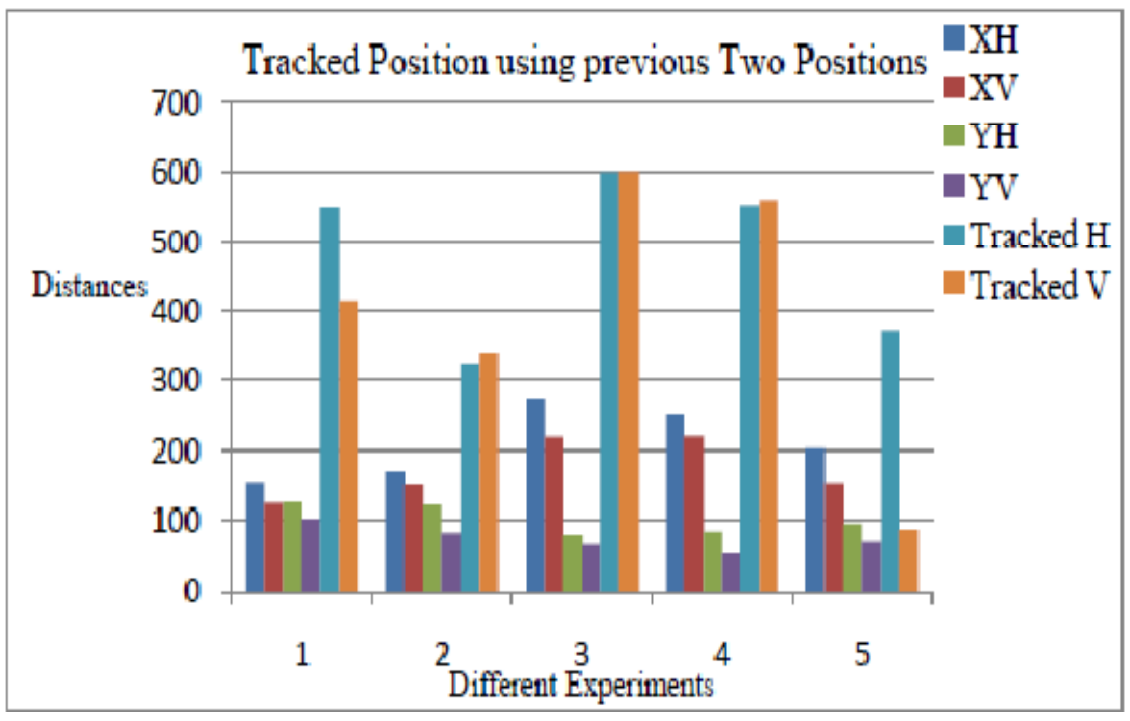

Fig. 16 Tracking Authenticated nodes

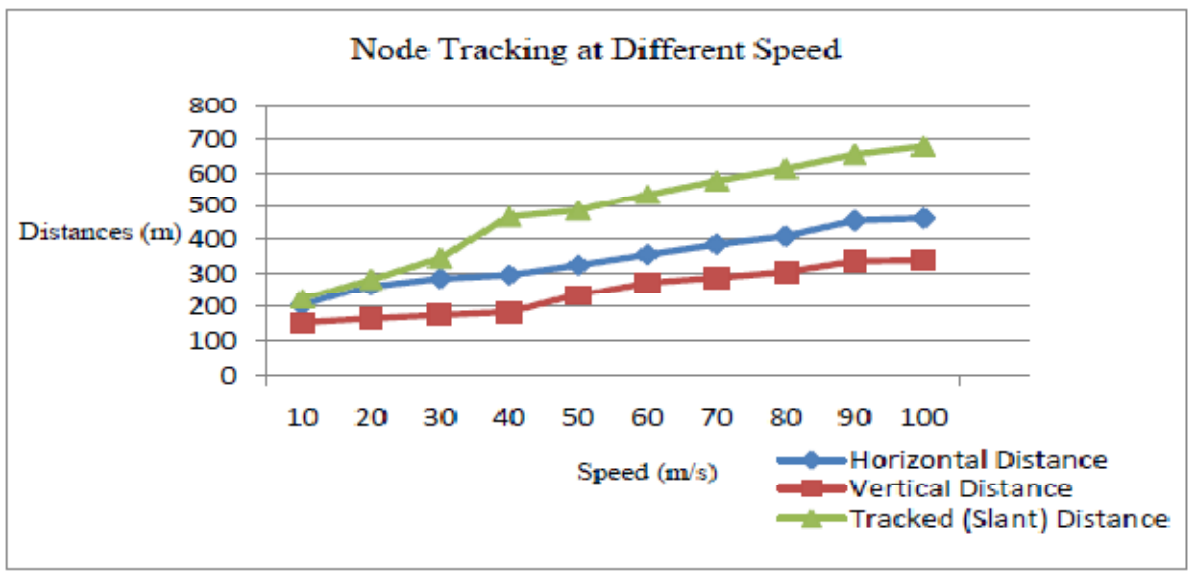

Fig. 17 Mobility versus Tracked distance

In the simulation of multi-lateration method, four known reference nodes A(4,70,20), B(42,26,70), C(112,26,38) and D(10,68,100), with reference to C(0,0,0) are selected near to malicious node. The average time difference of ToAs and ToDs are found 60,80 and 120 nano seconds between the reference and the target at an instance of time. For a different set of reference nodes, A(54,90,82), B(76,26,42), C(56,66,82) and D(80,54,48) near to the same malicious node at another instance of time, the average time difference of ToAs and ToDs are found to 100, 120 and 150 nano seconds. Then, using modified PL&T, the tracked location of the malicious node is determined at (59, 41) with respect to node A from the previous localization information of the malicious node and is shown in Fig. 18.

Different malicious nodes are tracked with 95 % detecting rate to the location replay and forgery deploying multilateration based PL&T with four known references and then modified PL&T depending upon their previous location set as{[(4,18),(38,32)],[(68,18),(64,32)],[(28,22),(40,35)],[(12,72),(36,52)],[(34,42),(54,38)]} computed from above multilateration method. The corresponding tracked location of malicious nodes are found {(59,41), (60,48), (61,59), (64,32), (11,48)} as shown in Fig. 19. The true PL&T of the malicious node matches with the computed positions fairly accurately using the proposed method.

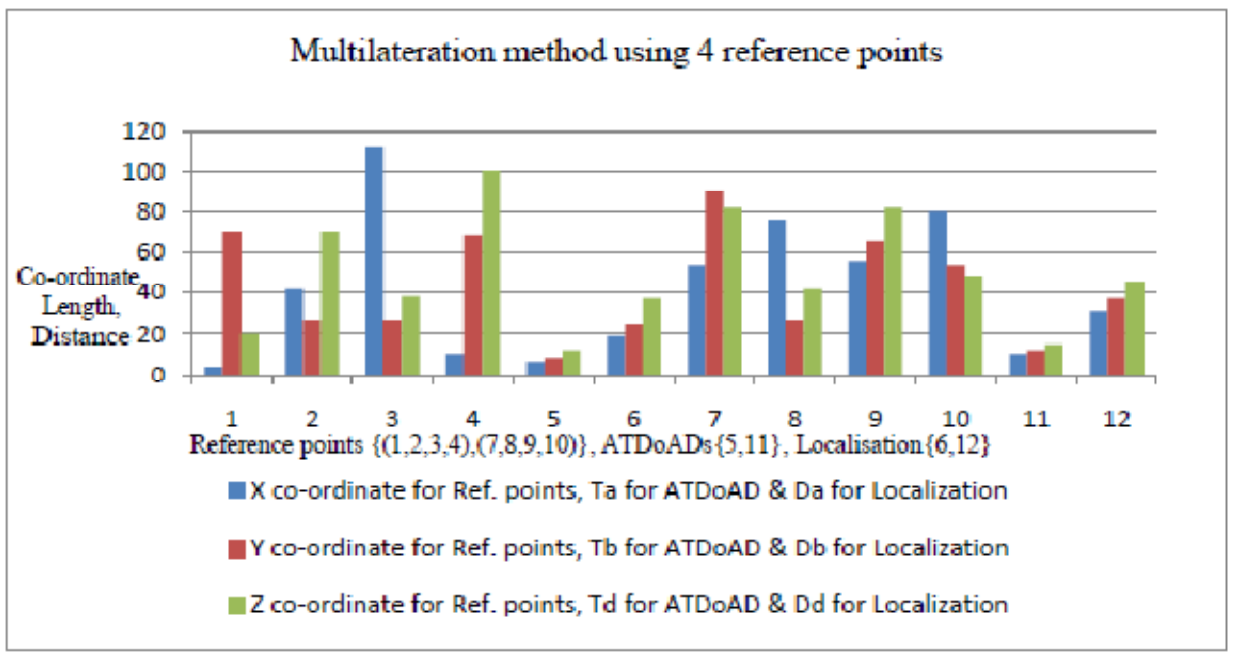

Fig. 18 Multi-lateration method for localizing malicious node

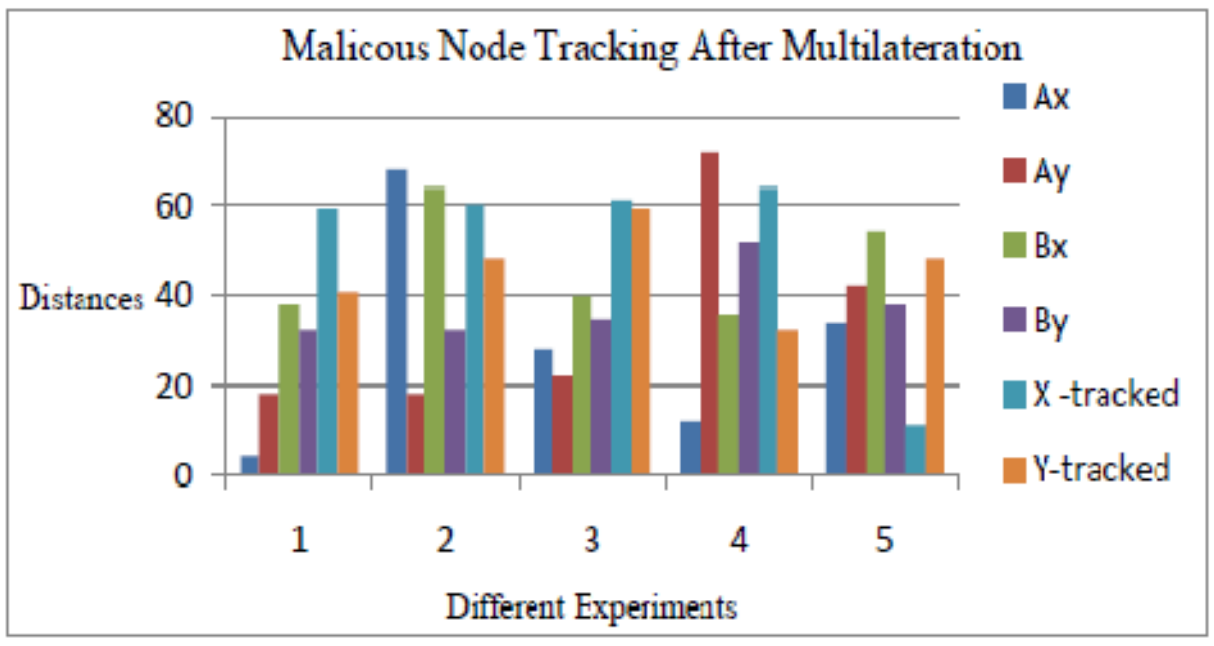

The tracking performance of triangulation (with modified PL&T) for authentic node and multi-lateration (with modified PL&T) for malicious node are illustrated in Fig. 20- Fig. 23. In the case of Triangulation, the average tracking error was found to be ~ 4.4163 m and a significantly higher error was there at sudden sharp turns in the trajectory shown in Fig. 20. The corresponding tracking error is illustrated in Fig. 21. The major cause of higher error in triangulation is that the reference nodes are too far from the authentic member nodes and they do not map location with respect to a single node for double cross check.

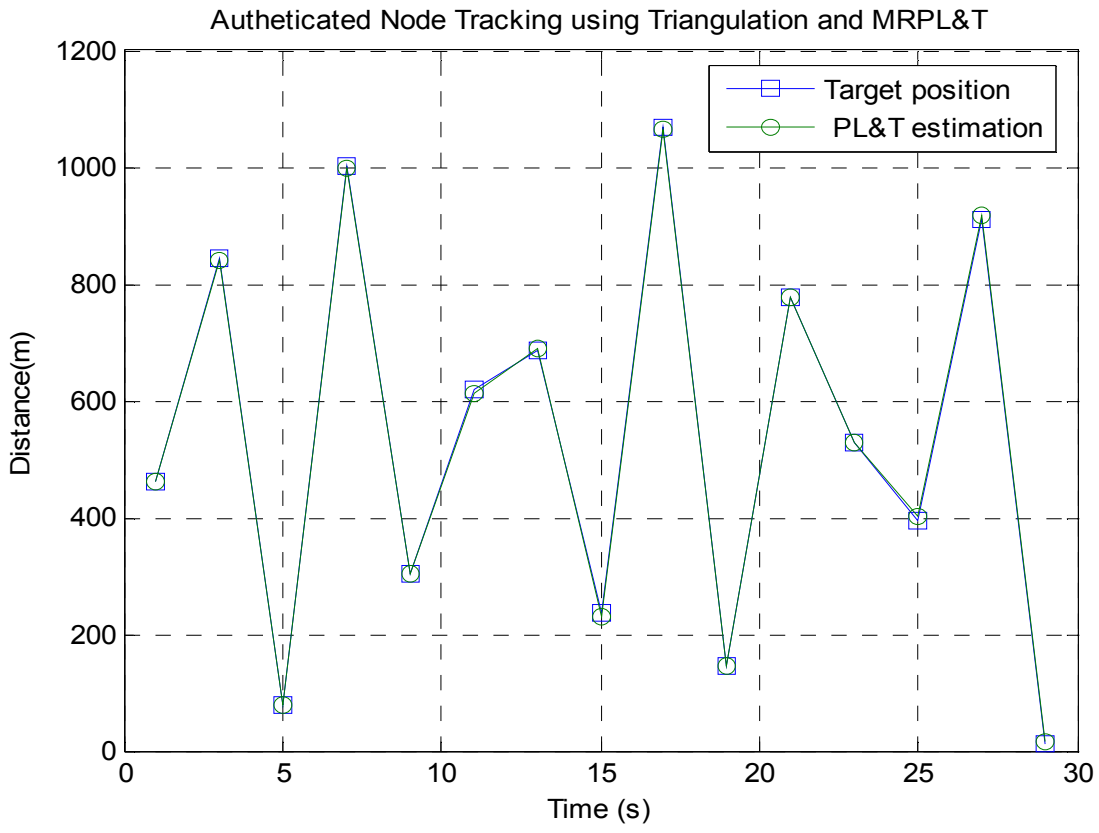

Fig. 20 Tracking Trajectory with Triangulation

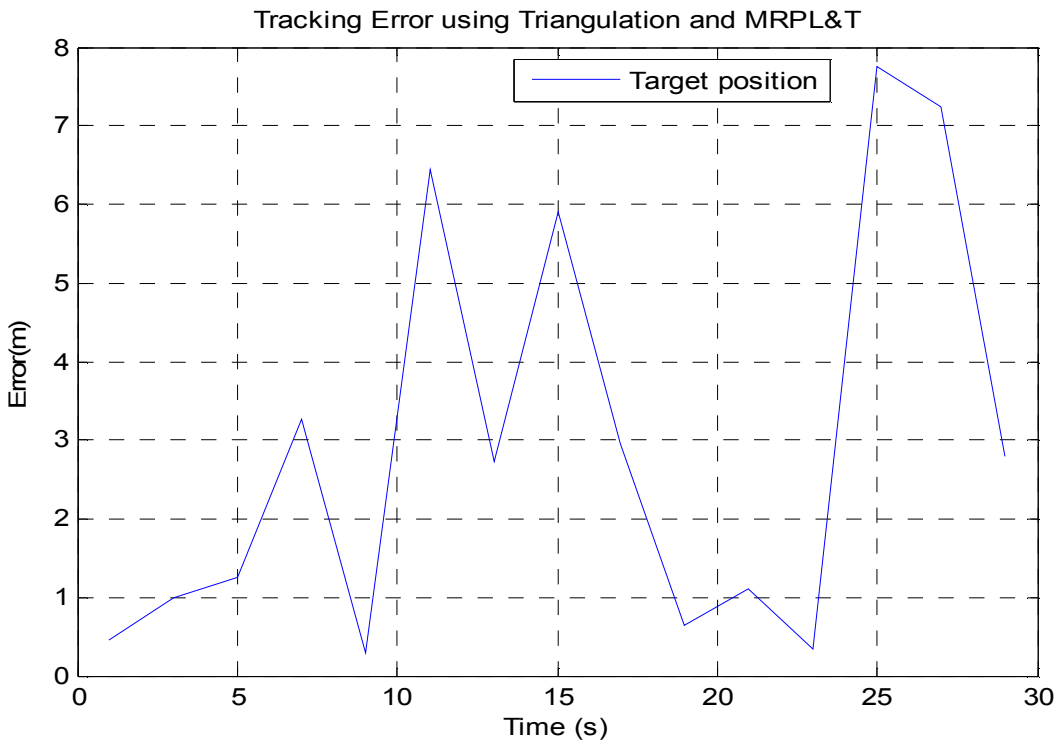

Fig. 21 Tracking Error in Triangulation

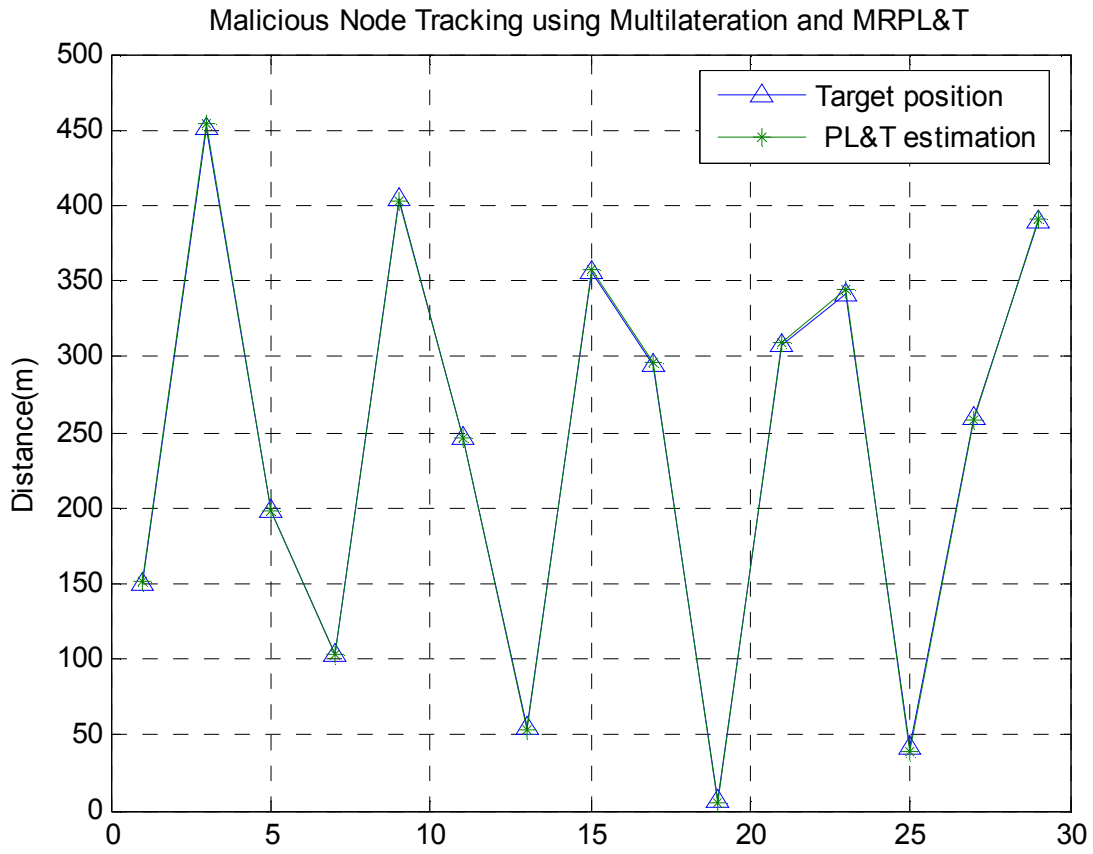

Fig. 22 Tracking Trajectory with Multi-lateration

In multi-lateralation, the average tracking error is found to be ~ 1.3 m and significantly higher error at sudden sharp turnings in trajectory as shown in Fig. 22. The corresponding tracking error is shown in Fig. 23. The major cause of lower error in multi-lateralation is that known reference nodes form triangulation which are in neighborhood of malicious node and they do mapping location with respect to a single node for double cross check.

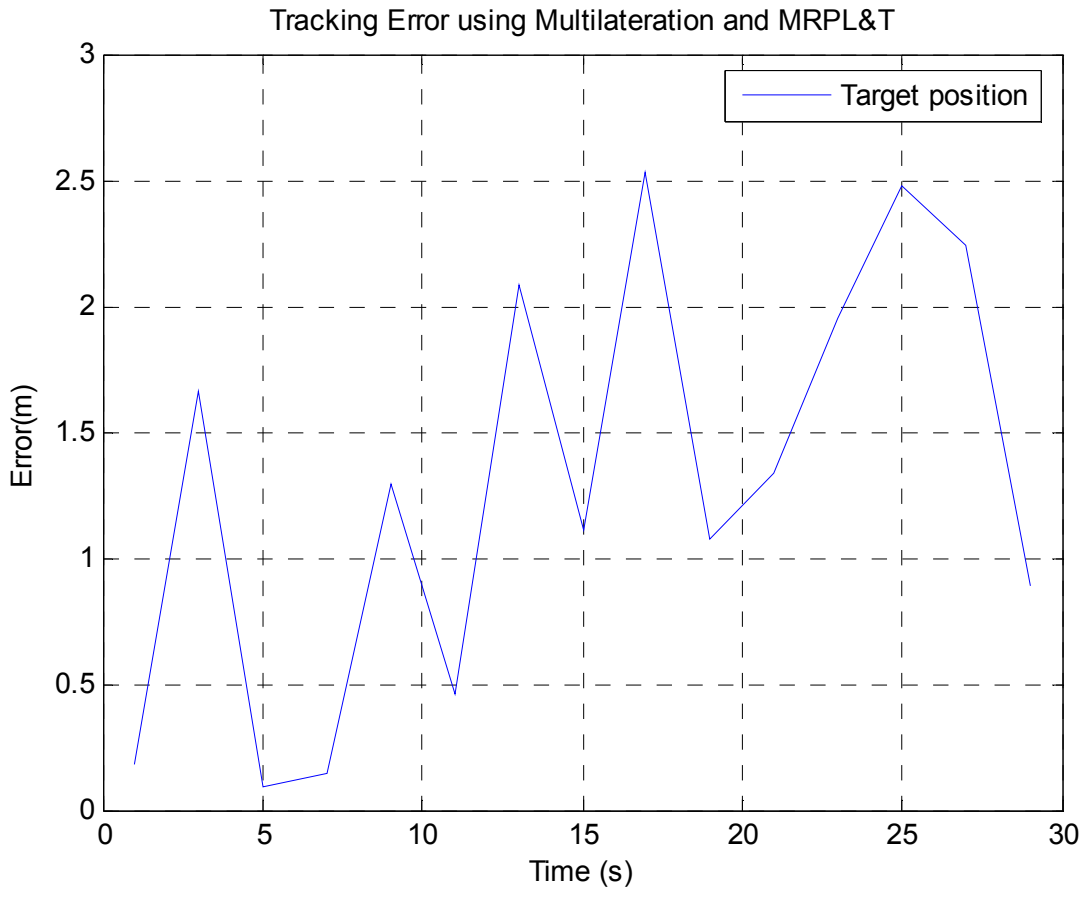

Fig. 23 Tracking Error in Multi-lateration

# 4. Conclusion

A robust, reliable and low complexity method of Distributed PL&T detection of malicious nodes in a cluster based MANET is proposed and its performance has been determined using simulation. This method includes algorithms for election of Cluster Heads, Registration Authorities to develop an overall architecture of the Dynamic Demilitarized Zone (DDMZ). This method prevents CA failures while reducing the security overhead traffic in clusters. The RAs operate as introducer to provide the public key and trust value to each requesting node about their target so that individual nodes can detect the malicious nodes in each cluster. This paper also presents the modified ToD and ToA based multi-lateration process which integrates the zone identification of the target node using adaptive beam forming which allows better geometry for PL&T operation. Tracking performance of multi-lateration with modified PL&T is found to be higher than when using the modified triangulation based PL&T. Our simulation show significant accuracy in the detection of malicious node and its PL&T data.

**ACKNOWLEDGMENT**

This research work is supported in part by the U.S. ARO under Cooperative Agreement W911NF-04-2-0054 and the National Science Foundation NSF 0931679. The views and conclusions contained in this document are those of the authors and should not be interpreted as representing the official policies, either expressed or implied, of the Army Research Office or the National Science Foundation or the U. S. Government.

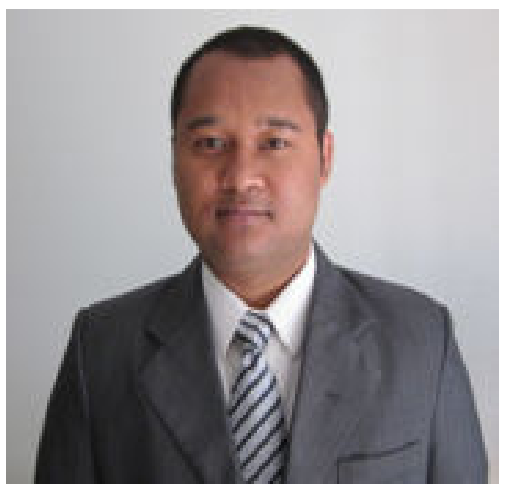

**Niraj Shakhakarmi** is a Ph.D student and working as a Research Assistant in the ARO Center for Battlefield Communications (CeBCom) Research, Department of Electrical and Computer Engineering, Prairie View A&M University since 2009. He received his B.E. degree in Computer Engineering from Kantipur Engineering College affiliated to Tribhuwan University, Nepal, in 2005 and M.Sc. in Information and Communications Engineering from Institute of Engineering, Pulchowk Campus, Tribhuwan University, Nepal, in 2007. His research interests are in the areas of cognitive radio networks, sensor networks, 4G networks and satellite networks. He is currently working on Position, Location & Tracking (PL&T), PL&T based security and Mobile Ad hoc Networks.

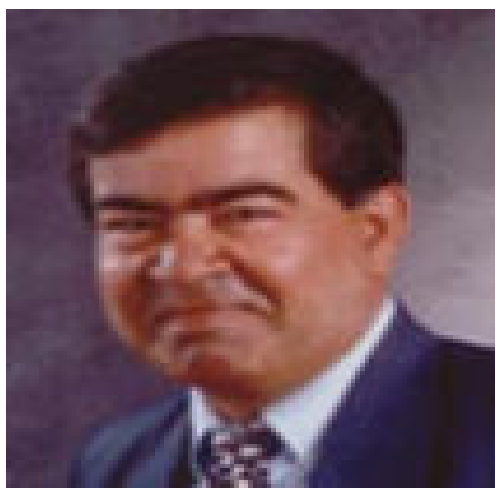

**Dhadesugoor R. Vaman** is Texas Instrument Endowed Chair Professor and Founding Director of ARO Center for Battlefield Communications (CeBCom) Research, ECE Department, Prairie View A&M University (PVAMU). He has more than 38 years of research experience in telecommunications and networking area. Currently, he has been working on the control based mobile ad hoc and sensor networks with emphasis on achieving bandwidth efficiency using KV transform coding; integrated power control, scheduling and routing in cluster based network architecture; QoS assurance for multi-service applications; and efficient network management.

Prior to joining PVAMU, Dr. Vaman was the CEO of Megaxess (now restructured as MXC) which developed a business ISP product to offer differentiated QoS assured multi-services with dynamic bandwidth management and successfully deployed in several ISPs. Prior to being a CEO, Dr. Vaman was a Professor of EECS and founding Director of Advanced Telecommunications Institute, Stevens Institute of Technology (1984-1998); Member, Technology Staff in COMSAT (Currently Lockheed Martin) Laboratories (1981-84) and Network Analysis Corporation (CONTEL) (1979-81); Research Associate in Communications Laboratory, The City College of New York (1974-79); and Systems Engineer in Space Applications Center (Indian Space Research Organization) (1971-1974). He was also the Chairman of IEEE 802.9 ISLAN Standards Committee and made numerous technical contributions and produced 4 standards. Dr. Vaman has published over 200 papers in journals and conferences; widely lectured nationally and internationally; has been a key note speaker in many IEEE and other conferences, and industry forums. He has received numerous awards and patents, and many of his innovations have been successfully transferred to industry for developing commercial products.